\documentclass[journal]{IEEEtran} 
\usepackage{xcolor,soul,framed} %,caption

\colorlet{shadecolor}{yellow}
\usepackage[pdftex]{graphicx}
\graphicspath{{../pdf/}{../jpeg/}}
\DeclareGraphicsExtensions{.pdf,.jpeg,.png,.eps}

\usepackage[cmex10]{amsmath}
\usepackage{array}
\usepackage{mdwmath}
\usepackage{mdwtab}
\usepackage{eqparbox}
\usepackage{url}

\newcommand{\E}{\varepsilon}
\newcommand{\B}{\beta}

\begin{document}
%\bstctlcite{IEEE:BSTcontrol}
    \title{Beamforming with metagratings at microwave frequencies: design procedure and experimental demonstration}
  \author{Vladislav~Popov,
      Fabrice~Boust,~\IEEEmembership{Member,~IEEE,}
      and~Shah~Nawaz~Burokur,~\IEEEmembership{Member,~IEEE}
     % <-this % stops a space

\thanks{© 20XX IEEE.  Personal use of this material is permitted.  Permission from IEEE must be obtained for all other uses, in any current or future media, including reprinting/republishing this material for advertising or promotional purposes, creating new collective works, for resale or redistribution to servers or lists, or reuse of any copyrighted component of this work in other works.}

  \thanks{V. Popov is with 	SONDRA, CentraleSup\'elec, Universit\'e Paris-Saclay, F-91190, Gif-sur-Yvette, France (e-mail: uladzislau.papou@centralesupelec.fr).}% <-this % stops a space
  \thanks{F. Boust is with 	SONDRA, CentraleSup\'elec, Universit\'e Paris-Saclay, F-91190, Gif-sur-Yvette, France
  and
  	DEMR, ONERA, Universit\'e Paris-Saclay, F-91123, Palaiseau, France
  (e-mail: fabrice.boust@onera.fr).}%
  \thanks{S. N. Burokur is with LEME, UPL, Univ Paris Nanterre, F92410, Ville d'Avray, France (e-mail: sburokur@parisnanterre.fr).}% <-this % stops a space
}

% The paper headers

% ====================================================================
\maketitle

\begin{abstract}
As opposed to metasurfaces,  metagratings represent themselves  sparse arrangements of scatterers.
Established rigorous analytical models allow metagratings to overcome performance of metasurfaces in beam steering applications while handling less degrees of freedom.
In this work we  deal with reflective metagratings that have only as few as one degree of freedom (represented by a reactively loaded thin wire)  per each propagating diffraction order.
We present a detailed design procedure and fabrication of three  experimental samples capable of establishing prescribed diffraction patterns.
The samples are experimentally studied in an anechoic  chamber dedicated to radar-cross-section bistatic measurements  and results are compared with three-dimension full wave numerical simulations.
We identify and analyze factors affecting operating frequency range of metagratings, suggest a strategy to increase the bandwidth.
\end{abstract}


\section{Introduction}

A diffraction grating, defined as a periodic optical structure with infinite extent in one direction diffracts waves incident on its surface~\cite{1}. 
Being imposed by the periodicity of a grating, which can be of the order of a free-space wavelength or greater, an incident wave is scattered as propagating diffraction orders only in certain directions. 
Concerning their applications, diffraction gratings have been widely used in laser resonators to tune and narrow lasing bandwidth~\cite{2,3}. 
Blazed or echelette gratings~\cite{4,5,6} capable of scattering an incident wave into a specific diffraction order have been applied in frequency-scanning reflector antennas~\cite{7,8,9,10} and for radar cross section (RCS) reduction~\cite{11,12} at microwave frequencies and in Littrow mount external cavity lasers in optics~\cite{13}. 
Classical blazed gratings are three-dimensional (3D) structures that generally take the form of right-angle sawtooths~\cite{14} and rectangular grooves~\cite{5}.

In the last few years, 2D metamaterials, also known as metasurfaces~\cite{15}, have  been applied to mimic blazed gratings functionality~\cite{16}. 
Metasurfaces representing themselves as very thin structures have been proposed as planar alternatives to metamaterials to exhibit light manipulation possibilities in various frequency domains, extending from microwave to visible frequencies. 
Local magnitude and phase of reflection and/or transmission coefficients of a metasurface can be controlled, and can thus be used to manipulate scattered wavefront of an incident beam. 
As such, metasurfaces have been used to perform functions including anomalous reflection and refraction~\cite{17,18,19,20,21,22}, deflection~\cite{23,24,25,26}, lensing~\cite{23,24,27,28,29,32}, thin-film cloaking~\cite{33,34,35}, coupling of propagating waves to surface waves~\cite{36,PhysRevB.97.115447}, optical vortex beams generation~\cite{37,38,39,40}, and holographic imaging ~\cite{41,42,43,44,45,46}, to name a few. 

\begin{figure}[tb]
\includegraphics[width=0.99\linewidth]{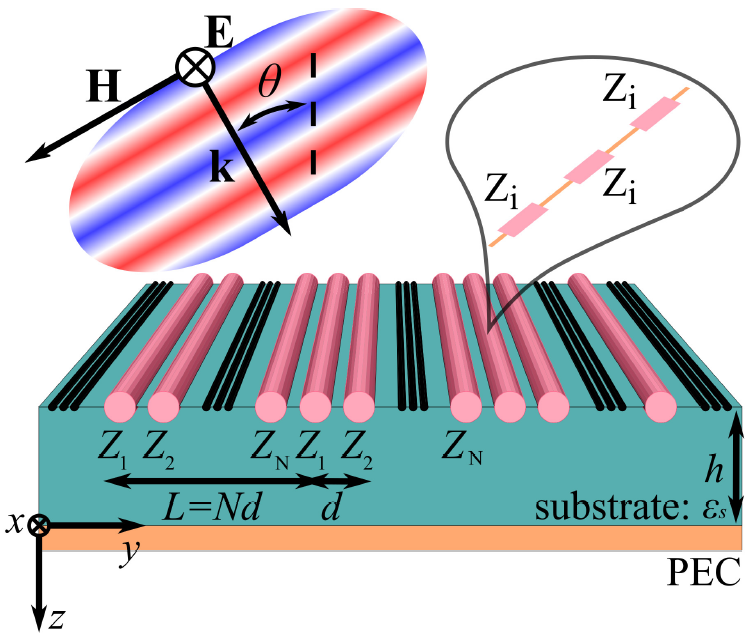}
\caption{\label{fig:1} Schematic diagram of the system under consideration: a periodic array of loaded thin wires (pink cylindrical lines) placed on PEC-backed dielectric substrate having permittivity $\E_s$ and thickness $h$. The array is excited by a TE-polarized plane wave incident at angle $\theta$.}
\end{figure}

Most of the metasurface-based wavefront manipulation geometries rely on the generalized laws of reflection and refraction presented in Ref.~\cite{17}. 
However several studies have shown that this approach suffer from low efficiency, particularly in configurations where extreme wave manipulation is considered (see, e.g., Refs.~\cite{Asadchy2016_SpatiallyDispMS,Alu2016_RHMS,49}). 
Moreover, implementation of field transformations into physical metasurface structures can reveal to be highly challenging and drawbacks concerning optimization time, design complexity of subwavelength periodically arranged resonant meta-atoms and material losses still exist. 

Very recently, the concept of metagratings evolved from classical diffraction gratings has been proposed as an interesting alternative to metasurfaces for boosting wavefront manipulation efficiency~\cite{Alu2017_mtg}. 
They are designed for diffraction engineering by cancelling a finite number of undesired propagating diffraction orders and allowing desired ones to radiate. 
In general, a metagrating is an array of scatterers (polarizable particles) separated by a distance of the order of the operating wavelength $\lambda$. 
The sparse arrangement of scatterers does not allow one describing metagratings in terms of local reflection and transmission coefficients  (or surface impedances) as metasurfaces.
In terms of meta-atoms, a metagrating consists of a limited number of meta-atoms in a supercell (period) compared to a metasurface which is composed of supercells incorporating numerous meta-atoms with subwavelength periodicity. 
Although metagratings can be considered as relatively simple systems in comparison to metasurfaces, functionalities such as perfect  anomalous reflection and perfect beam splitting have been demonstrated in Refs.~\cite{Alu2017_mtg,Epstein2017_mtg,Epstein2018_ieee_mtg,Epstein2019_mtg_exp}, where three propagating diffraction orders were considered at most and were handled by only two degrees of freedom.  
In Refs.~\cite{Popov2018,Popov2019_perfect}, the concept was generalized and the possibility to fully control an arbitrary number of propagating diffraction orders by means of a specific number of degrees of freedom was demonstrated.

Essentially metagratings can be understood by considering an example of  1D metagratings represented by a periodic array of supercells composed of $N$  thin  wires each. 
An incident wave excites polarization line currents in the wires resulting in the scattered field represented by Floquet-Bloch modes which are defined by the period $L$ of the array (i.e., the length of the supercell).
In particular, the diffraction angles of the propagating diffraction orders can be found via the grating formula: $L(\sin[\theta_m]-\sin[\theta_i])=m\lambda$, where $m$ represents the number of an order and $\theta_i$ is the incidence angle of an impinging  plane wave.
Furthermore, a line current is mathematically represented by the 2D Dirac delta function $\delta(y,z)$ that allows one to find the scattered field analytically, i.e., to know the complex amplitudes of all diffraction orders (propagating and nonpropagating).

In what follows, we deal with a particular configuration of 1D metagratings when thin wires are placed on the top of a metal-backed dielectric substrate as illustrated in Fig.~\ref{fig:1}. A plane-wave illumination is assumed and the wires interact only with the TE-polarized field.
As it was shown in the theoretical study~\cite{Popov2018}, the complex amplitudes $A_m^{\textup{TE}}$ of the electric field of the reflected plane waves are given by the following expression:
\begin{eqnarray}\label{eq:Am}
A_m^{\textup{TE}}&=&-\frac{k\eta}{2L}\frac{(1+R_m^{\textup{TE}})e^{j\B_mh}}{\B_m}\sum_{q=1}^{N}I_q e^{j\xi_m (q-1)d}\nonumber\\&+&\delta_{m0}R_0^{\textup{TE}}e^{2j\B_0h}
\end{eqnarray}
where $k$ and $\eta$ are respectively, the wavenumber and the characteristic impedance outside the substrate, $\xi_m=k\sin[\theta_i]+2\pi m/L$ and $\B_m=\sqrt{k^2-\xi_m^2}$ represent respectively, the tangential and normal components of wavevector of the plane waves, and $R_m^{\textup{TE}}$ is the corresponding Fresnel's reflection coefficient.
Equation~\eqref{eq:Am}  suggests that complex amplitudes of all $M$ propagating diffraction orders can be set arbitrarily if there are at least $N=M$ line currents $I_q$ in a supercell. 
Other parameters of the system, such as the parameters of the substrate and the distances between the line currents are assumed being fixed conversely to previously mentioned studies in Refs.~\cite{Alu2017_mtg,Epstein2017_mtg,Epstein2018_ieee_mtg,Epstein2019_mtg_exp}. 

Although here we focus on the TE polarization and reflective configuration of metagratings, the case of TM polarization can be studied similarly by means of duality relations (see, e.g., Refs.~\cite{felsen1994radiation,Popov2019_LPA}). 
The mathematical approach used in~\cite{Popov2018} to derive Eq.~\eqref{eq:Am} can be straightforwardly  generalized on transmissive-type metagratings, as the particular configuration studied in~\cite{Epstein2018_mtg_refraction}.

\begin{figure}[tb]
\includegraphics[width=0.99\linewidth]{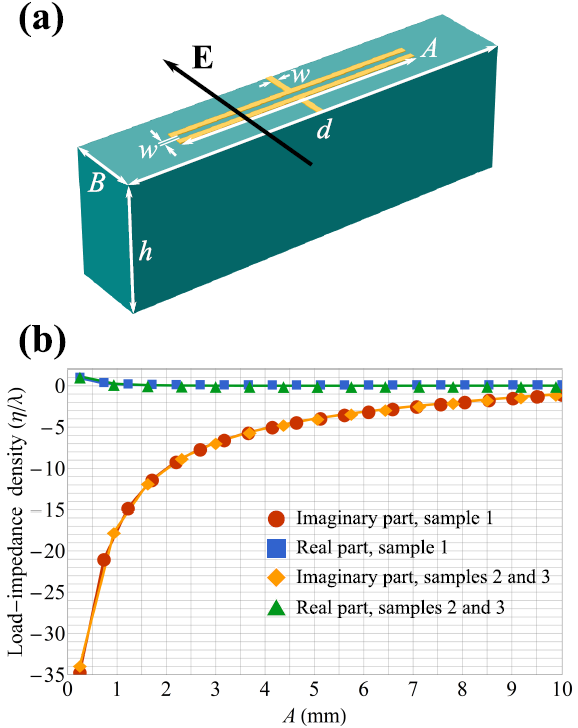}
\caption{\label{fig:2} (a) Schematic illustration of a capacitive unit cell: printed capacitance on top of a grounded dielectric substrate. (b) Load-impedance density of the printed capacitance extracted from specular reflection. $d\approx 11.6$ mm for the first sample and $d\approx 15.7$ mm in case of the second and third samples. Geometrical parameters are: $w=0.25$ mm, $B=3$ mm and $h=5$ mm and operating frequency is set to $10$ GHz.}
\end{figure}

In this work, based on the theoretical study of Ref.~\cite{Popov2018}, the design of simplified  metagratings composed of  the number of loaded wire as the considered number of propagating diffraction orders, is presented. 
The load-impedance densities of the wires are calculated and engineered from subwavelength wire elements. Measurements are performed on fabricated samples to experimentally validate the theoretical results .
The rest of the paper is organized as follows. 
In Section~\ref{sec:2} we provide the design methodology of the reflective-type metagratings. 
Section~\ref{sec:3} is devoted to the discussion of the experimental results and their comparison to simulation results.
In the same Section we discuss the mechanism behind the observed wide-band response of the proposed metagratings (see also Refs.~\cite{Epstein2018_ieee_mtg,Popov2018}).
Section~\ref{sec:4}   concludes the paper.

\begin{figure*}[tb]
\includegraphics[width=0.99\linewidth]{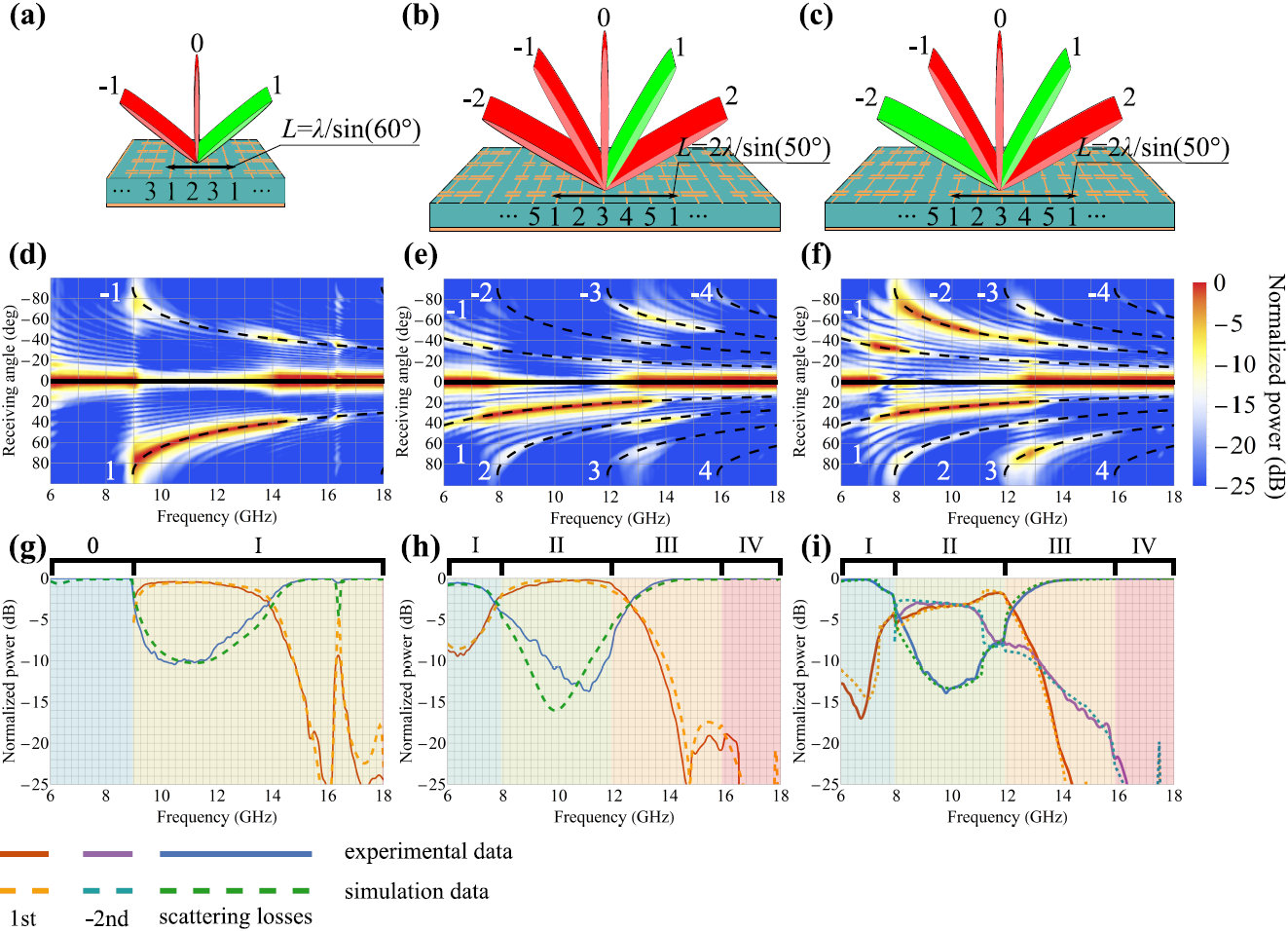}
\caption{\label{fig:3} (a)--(c) Schematics of prescribed diffraction patterns established by the three different designed metagratings with: (a) nonspecular reflection at an angle of $60^\circ$ with $N=3$ unit cells per period, (b) nonspecular reflection at an angle of $23^\circ$ with $N=5$ unit cells per period, and (c) equal excitation of the $-2^\textup{nd}$ and $+1^\textup{st}$ orders out of five diffraction orders, respectively. The green and red beams correspond to excited and suppressed diffraction orders, respectively. (d)--(f) Measurement results of the scattered power in the [6 GHz -- 18 GHz] frequency range. (g)--(i) Power management in the excited diffracting orders and scattering losses, the roman digits correspond to the highest propagating diffraction order in a given frequency range.}
\end{figure*}

\section{design procedure}
\label{sec:2}

In order to be able to control the diffraction pattern with a metagrating one has to carefully engineer it.
An appropriate dielectric substrate for a given frequency range is required. Its thickness $h$ and relative permittivity $\E_s$ should be carefully chosen in order to avoid excitation of waveguide modes~\cite{Popov2018}.
These waveguide modes are analog of surface plasmon polaritons responsible for well-known grating anomalies (or Wood's anomalies).
On the other hand, the presence of waveguide modes leads to divergence of certain Fresnel's reflection coefficients $R_m^{\textup{TE}}$ in Eq.~\eqref{eq:Am}, manifesting themselves in significant numerical errors.
Thus, in order to select a good substrate for a given metagrating's period $L$, one can plot the absolute value of the first few Fresnel's reflection coefficients corresponding to nonpropogating diffraction orders as a function of the substrate's parameters (thickness and permittivity) and avoid poles.
As a rule of thumb, a substrate with low permittivity and thickness of the order of $\lambda/(4\sqrt{\E_s})$ is a good candidate for the design of metagratings.

\begin{figure}[tb]
\includegraphics[width=0.99\linewidth]{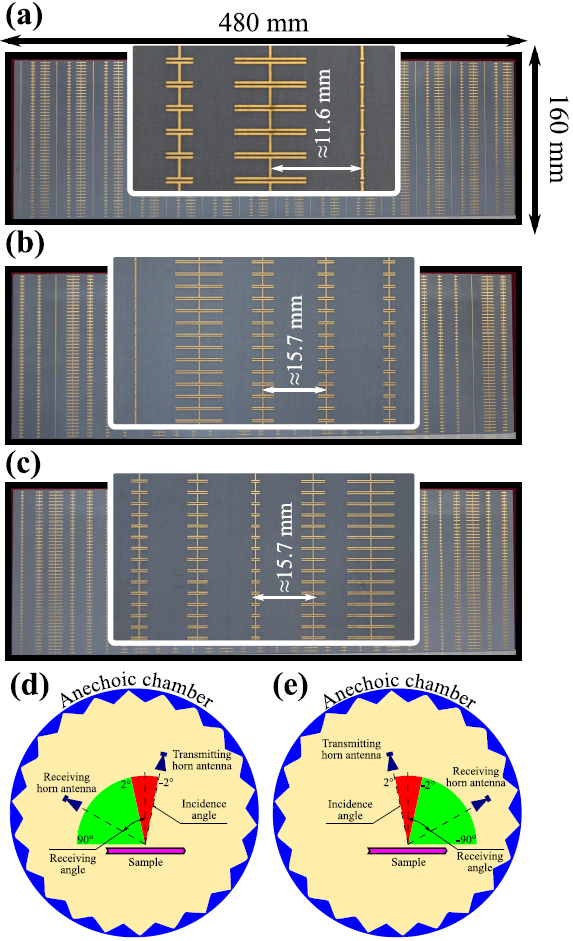}
\caption{\label{fig:4} (a)--(c) Photographies of the first, second and third samples (from top to bottom, respectively). (d), (e) Schematics of the experimental setup: to measure the scattering range of angles from $-90^\circ$ to $90^\circ$ the experiment is performed in the two steps illustrated by figures (d) and (e).}
\end{figure}

After selecting the correct substrate, the calculation of the characteristics of scatterers composing the metagrating has to be performed. 
Incident plane wave excites polarization line currents $I_q$ in  loaded thin wires that can be characterized by  load-impedance  $Z_q$ and input-impedance $Z_{in}$ densities. 
Each configuration of the diffraction pattern requires different set of load-impedance densities found from the Ohm's law:
\begin{equation}\label{eq:Z}
Z_qI_q=E_{q}-Z_{in}I_q-\sum_{p=1}^N Z_{qp}^{(m)}I_p.
\end{equation}
The right-hand side of Eq.~\eqref{eq:Z} represents the total electric field at the location of the $q^\textup{th}$ wire (including the self-action $Z_{in}I_q$). 
Thus, $E_q=(1+R_0^{\textup{TE}})\exp[j\B_0h-j\xi_0(q-1)d]$ is the external electric field created by the incident wave $e^{-j k \sin \theta y - j k \cos \theta z + j \omega t}$ reflected from the substrate and the sum $\sum_{p=1}^N Z_{qp}^{(m)}I_p$ takes into consideration the mutual interactions of the $q^\textup{th}$ wire with the rest of the wires (infinite number) and the grounded substrate.
The quantities $Z_{qp}^{(m)}$ are called as mutual-impedance densities.
Generally, load-impedance densities calculated from Eq.~\eqref{eq:Z} require engineering active and/or lossy response, i.e. $\Re[Z_q]\neq 0$. 
For instance,  in order to perform a large angle nonspecular reflection by means of a $N=M$ metagrating, one has to cancel two propagating diffraction orders out of $M=3$ available (as in the case of normal incidence).  
Then, the conditions $A_{-1}^{\textup{TE}}=0$ and $A_{0}^{\textup{TE}}=0$ leave one with only a single variable (being the phase of $A_1^{\textup{TE}}$) that cannot be used to satisfy three different equations
\begin{equation}\label{eq:reactive}
\Re\left[\left(E_q-\sum_{p=1}^N Z_{qp}^{(m)}I_p\right)I_q^*\right]-\Re[Z_{in}]|I_q|^2=0
\end{equation}
providing reactive load-impedance densities (the asterisk stands for the complex conjugate). 
Equation~\eqref{eq:Am} relates the complex amplitudes and currents, i.e. Eq.~\eqref{eq:reactive} can be rewritten in terms of $A_m$, with $m$ numbering  propagating diffraction orders.

%Mathematically, it means that power conservation conditions (set of $N$ algebraic equations obtained from Eq.~\eqref{eq:Z}) are not satisfied

In order to deal with  $N=M$ passive and lossless metagratings, equation~\eqref{eq:reactive} has to be satisfied. 
To that end, spurious scattering in undesired propagating diffraction orders has to be permitted. 
By introducing scattering losses, we sacrifice the efficiency for sake of design that would require only reactive elements.
An optimal configuration is achieved by numerically maximizing the power scattered in desired propagating diffraction orders while minimizing the left-hand-side of Eq.~\eqref{eq:reactive}.
It is worth to note that reflecting metasurfaces face the same difficulty (see for e.g., Refs.~\cite{Asadchy2016_SpatiallyDispMS,Alu2016_RHMS}) with notable exception of Refs.~\cite{Tretyakov2017_perfectAR,Epstein2016_AuxiliryFields}, which are rather special cases.
Generally, the efficiency of nonspecular reflection is used to evaluate the performance of conventional reflectarrays, i.e., efficiency decreases when the angle of nonspecular reflection increases.
However, highly efficient multichannel reflection can still be achieved as we demonstrate further.

\begin{table}[tb]
\caption{\label{tab:1}Parameters of the fabricated metagratings. The indexes correspond to the numbered unit cells in Figs.~\ref{fig:3} (a)--(c).}
\resizebox{0.48\textwidth}{!}{%
\begin{tabular}{|c|c|c|c|c|c|} 
\hline
Loads ($\eta/\lambda$) &$Z_1$ & $Z_2$ & $Z_3$ & $Z_4$ & $Z_5$  \\
 \hline
Sample 1 & $-j30.3$& $-j6.35$& $-j1.57$ & - & - \\
 \hline
Sample 2 &  $-j3.77$& $-j0.43$& $-j31.2$& $-j7.06$& $-j5.27$\\
 \hline
Sample 3 &  $-j3.75$& $-j4.84$& $j0.05$ &$-j2.94$& $-j8.86$ \\
\hline
 Arm's length (mm) & $A_1$ & $A_2$ & $A_3$ & $A_4$ & $A_5$\\ 
 \hline
Sample 1 & $0.37$ & $3.25$ & $8.70$ & - & -\\  
 \hline
Sample 2 &  $5.23$ & $11.2$ & $0.33$ & $2.91$ & $3.90$\\
 \hline
Sample 3 &  $5.25$ & $4.22$ & $12.3$ & $6.28$ & $2.27$\\
 \hline
\end{tabular}%
}
\end{table}

Once Eq.~\eqref{eq:reactive} is satisfied and corresponding complex amplitudes are found,  load-impedance densities are calculated from Eq.~\eqref{eq:Z} and implemented by wire elements engineered at subwavelength scale.
Although in a general case both capacitive and inductive loads might be required~\cite{Popov2019_perfect}, only capacitive elements are necessary in the examples considered further for an operating frequency set to $10$ GHz.
It is assumed that the samples would be fabricated by means of the conventional printed-circuit-board (PCB) technology.
Thus, thin wires are represented by metallic strips of width $w\ll \lambda$, thickness of $t_m=35$ $\mu$m, 
%the effective radius $r_0\approx w/4$  
and input-impedance density $Z_{in}=k\eta H_0^{(2)}(kw/4)/4$ as given in Ref.~\cite{tretyakov2003analytical}.
Capacitive loads are obtained by means of a microstrip printed capacitances, as illustrated in Fig.~\ref{fig:2} (a). 
Load-impedance density of printed capacitances can be approximated analytically using formulas for sheet impedance of a patch array~\cite{Tretyakov_patches_2008} as it is done in \cite{Epstein2017_mtg,Popov2018,Epstein2018_ieee_mtg,Popov2019_perfect}. 
Although the analytical model represents a simple tool for designing metagratings, it takes into account the mutual coupling with adjacent loaded wires via a phenomenological scaling parameter which is found by means of 3D full-wave simulations of an entire supercell and, thus, is not unique.
On the other hand, a recently developed simulation-based  approach~\cite{Popov2019_LPA} allows one to construct metagrating  unit cell by unit cell.
Instead of performing computations on a whole supercell, it deals with a single unit cell and takes analytically into account the interaction between adjacent wires to retrieve load-impedance density.
Additionally, simulation-based approaches are advantageous for being able to consider all practical aspects of meta-atoms, such as finite thickness of the metal cladding and conduction and dielectric losses.

\begin{figure}[tb]
\includegraphics[width=0.99\linewidth]{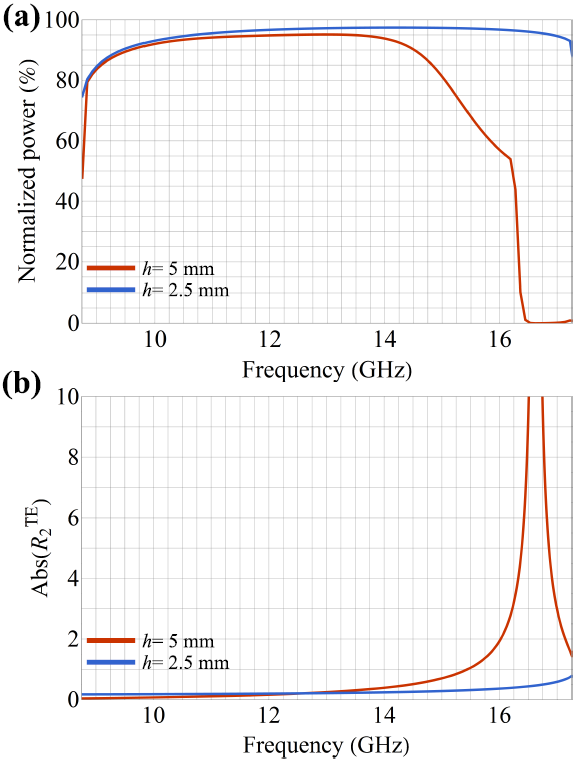}
\caption{\label{fig:5} (a) Computational results of the normalized power scattered by a reflective metagrating (having three reactive wires per period $L=\lambda/\sin(60^\circ)$) in the $+1^\textup{st}$ diffraction order vs. the frequency. Normally incident plane wave is assumed. Optimal reactive load-impedance densities are found at each frequency. (b) Absolute value of the Fresnel's reflection coefficient corresponding to the second (evanescent in the considered frequency range) diffraction order.}
\end{figure}

We design three experimental samples of metagratings to operate at $10$ GHz ($\lambda\approx 30$ mm) and we assume a normally incident  plane-wave illumination ($\theta=0$) for all three configurations. 
The functionalities of these three samples are schematically illustrated in Figs.~\ref{fig:3} (a)--(c).
The $h=5$ mm thick F4BM220 dielectric substrate having  permittivity $\E_s=2.2(1-j10^{-3})$ is selected as a good candidate for the proposed designs.
The first sample  deals with three diffraction orders maximizing the power scattered in the $+1^\textup{st}$ order and suppressing scattering in the the $-1^\textup{st}$ and $0^\textup{th}$ orders. Hence, it is composed of three unit cells per supercell, which has a length $L=\lambda/\sin(60^\circ)$ at $10$ GHz. This metagrating is able to achieve anomalous reflection at $60^\circ$ degrees.
The second and third samples each has five unit cells per supercell of length $L=2\lambda/\sin(50^\circ)$, which allows one to control five diffraction orders: $-2^\textup{nd}$, $-1^\textup{st}$, $0^\textup{th}$, $+1^\textup{st}$ and $+2^\textup{nd}$. 
The second sample maximizes the power scattered in the $+1^\textup{st}$ propagating diffraction order and thus performs small angle anomalous reflection, corresponding approximately to $23^\circ$ at $10$ GHz.
The third sample equally excites the $-2^\textup{nd}$ and $+1^\textup{st}$ orders while suppressing the three others.

On the basis of these specifications, the required load-impedance density are calculated by means of Eqs.~\eqref{eq:Am} -- \eqref{eq:reactive} and are presented in Table~\ref{tab:1}. Only two capacitive unit cells of length $d=\lambda/[3\sin(60^\circ)]$ and  $d=2\lambda/[5\sin(50^\circ)]$ need to be simulated for the design of the three metagratings.
A schematics of the unit cell is shown in Fig.~\ref{fig:2} (a).
In 3D full-wave simulations performed with COMSOL MULTIPHYSICS, periodic boundary conditions are applied to the side faces of the unit cell and the model is excited with a periodic port.
Parameters $w$ and $B$ are fixed to $0.25$ mm and $3$ mm, respectively.
The arm's length $A$ of the printed capacitance is used as a tuning parameter for the load-impedance density.
The load-impedance densities are first extracted from the $S_{11}$ parameter of the unit cell as detailed by the procedure in~\cite{Popov2019_LPA} and are plotted as function of $A$ in Fig.~\ref{fig:2} (b).
Although being built for two different parameters $d$, the two curves in Fig.~\ref{fig:2}(b) almost coincide.
It proofs that the analytical model used in Ref.~\cite{Popov2019_LPA} to take into account the interaction between adjacent wires and the substrate, allows one to obtain the load-impedance density of a wire itself and not of a corresponding  array.
Eventually, the load-impedance densities are used to tailor the geometrical parameters of the microstrip printed capacitances listed in Table~\ref{tab:1}.
Photographies of the fabricated samples are displayed in Fig.~\ref{fig:4} (a)--(c) and their physical size is approximately $480$ mm ($y$-direction) by $160$ mm ($x$-direction).

\section{Experimental results}
\label{sec:3}

In this section we demonstrate experimentally the control of diffraction patterns with the proposed and fabricated metagrating designs.  
The  samples are tested in an anechoic chamber dedicated to radar-cross-section bistatic measurements, where transmitting and receiving horn antennas are mounted on a circular track of $5$ m radius. 
A schematic representation of the experimental setup is shown in Figs.~\ref{fig:4} (d) and (e).
In the current experiments, the transmitter is fixed and the receiver moves with $1^\circ$ step and the minimum angle between the transmitter and receiver for the scanning is $4^\circ$.
In order to be able to measure the specular reflection, the transmitter is fixed at $\mp2^\circ$.
Thus, the experiments are conducted in two steps: when the transmitter is fixed at $\mp2^\circ$, the receiver moves form $\pm2^\circ$ to $\pm90^\circ$, as it is clearly illustrated in  Figs.~\ref{fig:4} (d) and (e).

Figures~\ref{fig:3}(d)--(f) visualize  angle measurements of the scattered power in the frequency range spanning from $6$ to $18$ GHz.
It is clearly observed that the positions of the main lobes (corresponding to diffraction orders) are in perfect agreement with the results given by the grating formula $\theta_m=\sin^{-1}(m c/(\nu L)+\sin[\theta_i])$ (represented by black dashed curves). Here, $c$  is the speed of light in vacuum and $\nu$ is the frequency.
However, the spectrum of waves scattered from a finite-size sample in the far-field is  much more complex than just a few plane waves representing propagating diffraction orders.
Thus, in order to estimate the performance of the samples we execute next steps following Ref.~\cite{Epstein2019_mtg_exp}.
In the first place, we localize each diffraction order between the angles $\theta_m^{(1)}$ and $\theta_m^{(2)}$ which correspond to $3$ dB of the power attenuation with respect to the maximum power of the lobe.
The maximum is found near the angle $\theta_m=\sin^{-1}(m c/(\nu L)+\sin[\theta_i])$.
Finally, the normalized power $f_m(\nu)$ scattered in a given diffraction order $m$ at the frequency $\nu$ is estimated by means of the following integral formula
\begin{equation}\label{eq:perf}
f_m(\nu)=\frac{\int_{\theta_{m}^{(1)}}^{\theta_{m}^{(2)}} P(\nu,\theta)d\theta}{\sum_{m}\int_{\theta_{m}^{(1)}}^{\theta_{m}^{(2)}} P(\nu,\theta)d\theta},
\end{equation}
where $P(\nu,\theta)$ is the absolute power scattered in the receiving angle $\theta$ at the frequency $\nu$.
The summation in the denominator is performed over all propagating diffraction orders at the frequency $\nu$.
Figures~\ref{fig:3}(g)--(i) show the performance of the experimental samples (solid curves obtained by means of Eq.~\eqref{eq:perf}) as function of the frequency, scattering losses represent the power scattered in undesired diffraction orders.
The dashed curves demonstrate the results obtained from 3D full-wave simulations (a supercell with imposed periodic boundary conditions and excited by a periodic port).
By comparing the solid and dashed curves, one can observe a good agreement between the experimental and simulation results.

Although the samples were designed to operate at a single frequency ($10$ GHz), it is seen that the scattering losses remain low in a wide range of frequencies.
One of the most important factors affecting an operating frequency range is the frequency response of unit cells. Resonant elements, in a general manner, significantly decrease an operating frequency range (see, e.g., Refs.~\cite{Popov2019_perfect,Popov2019_LPA}).
As demonstrated by Fig.~\ref{fig:2}(b), unit cells used to construct experimental samples do not exhibit resonances at $10$ GHz.
Since the designed metagratings possess a number of degrees of freedom equal to the number of propagating diffraction orders, it is expected that the scattering losses increase when approaching frequencies where the number of propagating diffraction orders changes (corresponding to different areas in Figs.~\ref{fig:3}(g)--(i) labeled with roman digits).
While it is the case for the second and third samples, the performance of the first one decreases far before the appearance of the second propagating diffraction orders, see Figs.~\ref{fig:3}(g)--(i).
It unveils yet another crucial factor influencing an operating frequency range: excitation of waveguide modes discussed in the very beginning of Section~\ref{sec:2}.
Although we avoid waveguide modes around the design frequency of $10$ GHz, they may appear at lower or higher frequencies and this is exactly what happens with the first sample, as we further present in Fig.~\ref{fig:5}.
A waveguide mode is excited at the frequency when the Fresnel's reflection coefficient $R_2^{\textup{TE}}$ diverges and leads to drastic decrease of the performance of the metagrating, as it can be clearly observed in Fig.~\ref{fig:5} when comparing two different thicknesses of the dielectric substrate.
In the experimental and simulation data the waveguide mode manifests itself in the resonance observed around  $16.4$ GHz, see Figs.~\ref{fig:3}(d) and (g).
Figure~\ref{fig:5}(a) presents the computational results of maximizing the power of a normally incident plane wave coupled to the $+1^\textup{st}$ propagating diffraction order in the three unit cells per period metagrating, assuming purely reactive load-impedance densities. As demonstrated, the excitation of the waveguide mode can be suppressed by choosing a thinner substrate (for e.g. $2.5$ mm instead of $5$ mm) which enables restoring the performance over the entire range of frequencies where there are three propagating diffraction orders (see blue curves in Fig.~\ref{fig:5}).

\section{conclusion}
\label{sec:4}

To conclude, we have described in details the design procedure of reflective metagratings and tested three experimental samples able of establishing prescribed diffraction pattern in a wide frequency range.
The experimental results have demonstrated a good agreement with 3D full-wave simulations.
Thus, we have experimentally verified the concept of metagratings for controlling multiple beams with as few as one  degree of freedom (represented by a reactive load) per a propagating diffraction order.
We have identified the main factors affecting the operating frequency range of metagratings which should facilitate the development of wide-band beamforming devices in the future.
%Sparse arrangement of elements in metagratings

\section*{acknowledgements}

The authors acknowledge the help of Anil Cheraly  (ONERA) in conducting the experiments.

%\bibliography{bib}
\bibliographystyle{IEEEtran}
\bibliography{IEEEabrv,bib}
\end{document}